\documentclass[useAMS,usenatbib,usegraphicx]{mn2e}
\usepackage{epsfig}
\usepackage{amsmath} %need for line break in equ 3
\usepackage{rotating}           % for sideways tables/figures
\usepackage{color}     
\usepackage{graphicx}
\usepackage{times}
\usepackage{upgreek} % for up $\uptau$ 
\usepackage{amssymb} % for \checkmark

\def\kms{km ${\rm s}^{-1}$}

\def\scm  {$\hbox{{\rm cm}}^{-2}$}    %cm-2
\def \AL {$\alpha $}     %  gr. alpha
\def \HI {H{\sc \,i}}
\def \WpHz {W Hz$^{-1}$}
\def\lapp{\ifmmode\stackrel{<}{_{\sim}}\else$\stackrel{<}{_{\sim}}$\fi}
\def\gapp{\ifmmode\stackrel{>}{_{\sim}}\else$\stackrel{>}{_{\sim}}$\fi}
\title[Mid-infrared properties of active galaxies]{The mid-infrared properties and gas content of active galaxies over large look-back times}
\author[S. J. Curran,  et al.]{S. J. Curran\thanks{Stephen.Curran@vuw.ac.nz} and   S. W. Duchesne \\
School of Chemical and Physical Sciences, Victoria University of Wellington, PO Box 600, Wellington 6140, New Zealand\\
}

\begin{document}

\date{Accepted ---. Received ---; in original form ---}

\pagerange{\pageref{firstpage}--\pageref{lastpage}} \pubyear{2018}

\maketitle

\label{firstpage}

\begin{abstract}
  Upon an expansion of all of the searches for redshifted \HI\ 21-cm absorption ($0.0021\leq z \leq5.19$), we update recent
  results regarding the detection of 21-cm in the non-local Universe. Specifically, we confirm that photo-ionisation of
  the gas is the mostly likely cause of the low detection rate at high redshift, in addition to finding that at
  $z\lapp0.1$ there may also be a decrease in the detection rate, which we suggest is due to the dilution of the
  absorption strength by 21-cm emission. By assuming that associated and intervening absorbers have similar cosmological
  mass densities, we find evidence that the spin temperature of the gas evolves with redshift, consistent with heating
  by ultra-violet photons. From the near--infrared ($\lambda=3.4$, 4.6 and 12~$\mu$m) colours, we see that radio
  galaxies become more quasar-like in their activity with increasing redshift.  We also find that the non-detection of
  21-cm absorption at high redshift is not likely to be due to the selection of gas-poor ellipticals, in addition to a
  strong correlation between the ionising photon rate and the $[3.4] - [4.6]$ colour, indicating that the UV photons
  arise from AGN activity. Like previous studies, we find a correlation between the detection of 21-cm absorption and
  the $[4.6] - [12]$ colour, which is a tracer of star-forming activity.  However, this only applies at the lowest
  redshifts ($z\lapp0.1$), the range considered by the other studies.
\end{abstract} 


\section{Introduction}
\label{intro}
 
Cool, neutral gas, the reservoir for star formation throughout the Universe, is traced by the absorption of the
$\lambda=21$-cm continuum radiation by neutral hydrogen (\HI). This can be detected in either the gas in galaxies {\em
  intervening} the sight-line to a background continuum source (quasar) or by the gas {\em associated} with the radio
galaxy or quasar itself.  The conditions to which the cool gas is exposed differs in each of these cases, with the
intervening absorption arising in quiescent galaxies, which may trace the star-formation history
(\citealt{cur17,cur17a}).  The associated absorption arises within the host galaxies of active galactic nuclei (AGN)
and, since the total neutral hydrogen column density, $N_\text{\HI}$, is generally not known, the spin temperature of
the gas (degenerate with the covering factor, see \citealt{cur12}) cannot be determined, which does not allow a measure
of how the fraction of cool gas evolves with redshift.  There has, nevertheless, been some recent advances in
understanding the nature of the neutral gas in the associated absorbers:
\begin{enumerate}
\item The paucity of detection of 21-cm absorption at $z\gapp1$ may be due to the optical selection of sources,
  leading to ultra-violet luminosities ($L_{\rm UV}\gapp10^{23}$ \WpHz) sufficient to ionise all of the neutral gas
  \citep{cww+08,cw12}.
\item The similar detection rate for both type-1 and type-2 AGN below $L_{\rm UV}\sim10^{23}$ \WpHz\ indicates 
  that the absorption does not arise predominately in the sub-pc torus (e.g. \citealt{pvtc99,gss+06}), 
but in the large-scale galactic disk \citep{cw10}.
\item The higher detection rate in compact objects (e.g. \citealt{vpt+03,gmo14}) may be due to the fact that these
``young'' sources have lower UV luminosities than the extended objects \citep{cw10,ace+12}.
\item The absorption strength is correlated with both the visible--near-infrared ($V-K$, \citealt{cw10}) and
  blue--near-infrared ($B-K$, \citealt{cwa+17}) colours, indicating that the reddening is caused by intrinsic dust,
  which shields the cool, neutral gas from the ambient UV field.
\item The anti-correlation between the strength of the absorption and the size of the radio source
  (e.g. \citealt{pcv03,css11}) can be explained by the effect of geometry on the coverage of the continuum source
  \citep{cag+13}. This is also evident in the increase in detection rate with turnover frequency for the gigahertz peaked
spectrum sources \citep{cwa+17}.
\item The molecular gas, detected in many instances through CO emission at $z\gapp3$, is remote from the continuum
  source (e.g. \citealt{dnm+03,efr+13}), and thus not subject to the same ionising flux as the neutral gas
  \citep{caw+16}.
\item There is a strong relationship between 21-cm and soft X-ray absorption, each of which is consistent with the presence of dense neutral gas \citep{oms+10, omd+17,gas+17,mas+17}.
\item The detection of 21-cm absorption is related to the mid-infrared colours of the source. Specifically, a detection
is more likely at a higher 4.6--12 $\mu$m colour, which is a tracer of star forming activity \citep{cs17,gas+17}.
\end{enumerate}
This latter item has only been investigated at low redshift ($z\lapp0.2$) and, given its potential importance in future
surveys for \HI\ 21-cm absorption with the Square Kilometre Array \citep{msc+15}, should be expanded to include all of
the redshifted 21-cm searches, which we do in this paper. Since there has been some recent discussion questioning the
importance of the UV luminosity, in Sect.~\ref{epf} we revisit this argument in context of the radio luminosity, which
is also advocated as  being a crucial factor in the detection of 21-cm absorption. In Sect.~\ref{eng}, we discuss the variation in
the detection rate at various redshifts. In Sect.~\ref{nic}, we examine any
correlation between the mid-infrared colours and the detection of 21-cm absorption, in addition to investigating how
this is related to the ionising UV continuum.  Finally, in Sect.~\ref{sum} we summarise our conclusions.

\section{The effect of the background continuum}
\label{epf}

\subsection{The updated sample}

\begin{figure}
\centering 
\includegraphics[angle=270,scale=0.52]{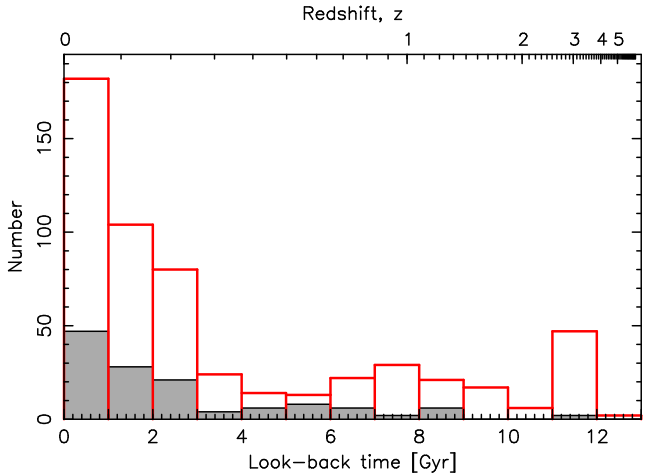}
\caption{The distribution of the 21-cm detections (shaded histogram) and non-detections (unshaded) with look-back time.
  Throughout the paper we use a standard $\Lambda$ cosmology with $H_{0}=71$~km~s$^{-1}$~Mpc$^{-1}$, $\Omega_{\rm
    matter}=0.27$ and $\Omega_{\Lambda}=0.73$.}
\label{LBT-histo}
\end{figure}

Previously (e.g. \citealt{chj+17}), we only considered the $z\geq0.1$ \HI\ 21-cm absorption searches, giving 311 sources
for which limits could be obtained (i.e. of sufficiently high signal-to-noise and not dominated by radio frequency
interference). We now expand the sample by including all of the published redshifted searches (Fig. \ref{LBT-histo}), giving a sample of 689
sources (comprising 129 detections and 560 limits), spanning redshifts of $0.0021\leq z \leq 5.19$ (look-back times of 0.03 to
12.5 Gyr).
These have been compiled from
\citet{dsm85,mir89,vke+89,ubc91,cps92,cmr+98,cwh+07,mcm98,ptc99,ptf+00,rdpm99,mot+01,ida03,vpt+03,cwm+06,cww+08,cwm+10,cwwa11,cwsb12,cwt+12,caw+16,chj+17,cwa+17,gss+06,omd06,kpec09,ems+10,ssm+10,css11,cgs13,ace+12,asm+15,ysdh12,ysd+16,gmmo14,sgmv15,akk16,akp+17,gas+17,mas+17,
  mmo+17,omd+17, gdb+15}\footnote{\citet{gdb+15}, which is still in submission, reports 0 new detections of \HI\ 21-cm
  absorption out of 89 new searches over $0.02 < z < 3.8$ (see \citealt{gd11}).} and will be available in a
forthcoming on-line catalogue (Moss et al., in prep).

As described in \citet{cwsb12}, we obtain the photometry of each source from the NASA/IPAC Extragalactic Database (NED),
the Wide-Field Infrared Survey Explorer (WISE, \citealt{wem+10}), Two Micron All Sky Survey (2MASS, \citealt{scs+06})
and the Galaxy Evolution Explorer (GALEX data release GR6/7)\footnote{http://galex.stsci.edu/GR6/\#mission} databases,
which are then shifted into the rest-frame of the source and converted to a luminosity upon correction for Galactic extinction
\citep{sfd98}.

\subsection{Ultra-violet}
\label{ciul}

From a survey for \HI\ 21-cm absorption at high redshift, \citet{cww+08} obtained zero detections out of the ten
$z\gapp3$ sources searched. Upon a detailed investigation of the source photometries, they suggested that the high
redshifts introduced a selection effect, where the faint optical targets ($B\gapp19$) were in fact UV bright in the
source rest-frame, thus ionising/exciting the gas to below the detection limit.  Although this ``critical UV
luminosity'' of $L_{\rm UV}\sim10^{23}$ \WpHz\ has not yet been falsified by several consequent surveys
(\citealt{cwm+10,cwsb12,cwt+12,caw+16,chj+17,cwa+17,ace+12,gmmo14,akk16,akp+17,ak17,gdb+15}), \citet{akk16} and
\citet{ak17} suggest that other effects may be responsible for the non-detections. Namely, gas excitation by the
incident radio continuum or some unspecified redshift evolution.

\begin{figure*}
\centering 
\includegraphics[angle=270,scale=0.59]{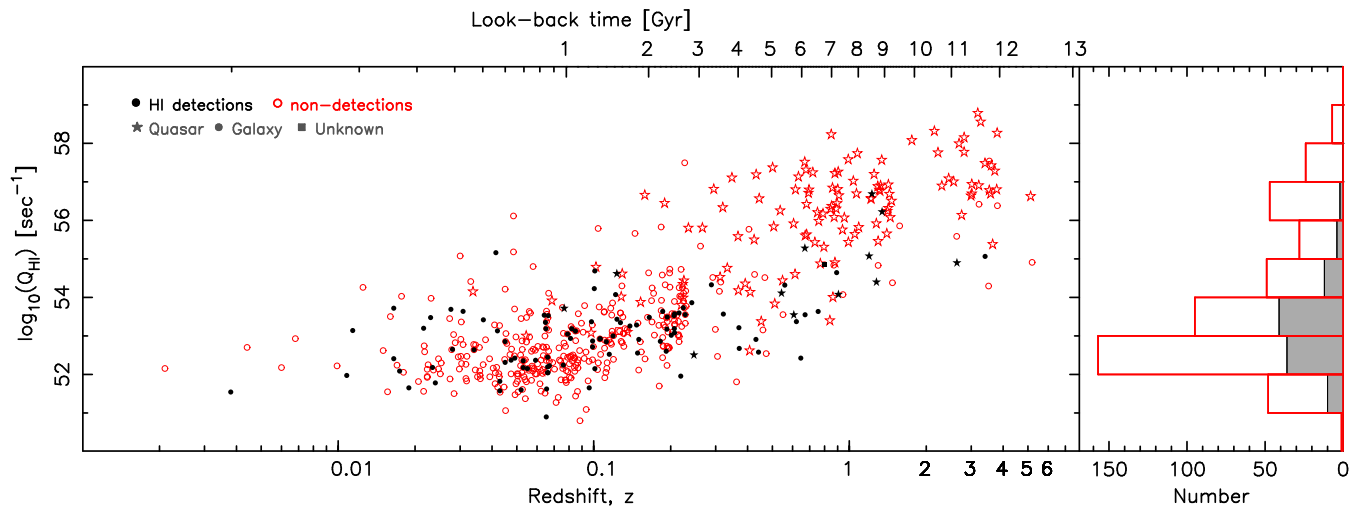}
\caption{The ionising ($\lambda \leq912$ \AA) photon rate versus redshift for the \HI\ 21-cm absorption searches.  The
  filled circles/histogram represent the detections and the unfilled circles/histogram the non-detections, with the
  shapes representing the classification as given by the NASA/IPAC Extragalactic Database (NED).}
\label{Q-z}
\end{figure*}

\begin{figure}
\centering 
\includegraphics[angle=270,scale=0.52]{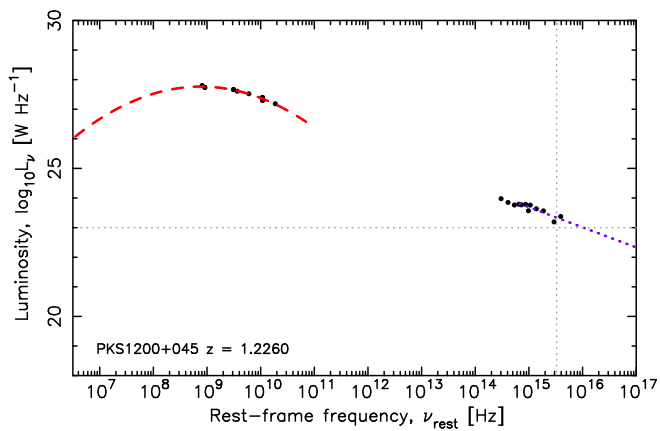}
\caption{The available photometry of PKS\,1200+045 overlaid by fits to the radio and optical/UV points (see
  \citealt{cwsb12}). The vertical dotted line signifies a rest-frame frequency of $3.29\times10^{15}$ Hz ($\lambda =
  912$~\AA) and the horizontal a $\lambda = 912$~\AA\ luminosity of $L_{\rm UV}=10^{23}$ \WpHz.}
\label{1200}
\end{figure}

Before addressing this, it should be noted that \citeauthor{akk16} use the $\lambda=912$~\AA\ monochromatic luminosity
\citep{cww+08}, when in fact it is the {\em total} bolometric luminosity at $\lambda \leq912$~\AA, which should be used
\citep{cw12}.  The ionising photon rate, $Q_\text{\HI}\equiv \int^{\infty}_{\nu}({L_{\nu}}/{h\nu})\,d{\nu}$, is obtained from a
power-law fit to the extinction corrected blue/UV ($\nu\gapp10^{15}$~Hz) photometry, which we show for all of the sources which
have sufficient data in Fig.~\ref{Q-z}.
The highest ionising photon rate at which \HI\ has been detected
($Q_\text{\HI}=4.8_{-1.9}^{+3.2}\times10^{56}$~sec$^{-1}$) is now due to PKS\,1200+045 \citep{ak17}.\footnote{From
  $L_{\rm UV}\approx Q_\text{\HI}h$, this gives a monochromatic luminosity of $L_{\rm UV}\approx 3\times10^{23}$ \WpHz.}
This value is quite unreliable since the SED has only one value above the ionising frequency (Fig.~\ref{1200}) and does
not include a break in the power-law.\footnote{The UV continuum in Quasi-Stellar Objects (QSOs) is seen to follow a
  broken power law, with a break close to $\lambda =1200$~\AA\ and spectral indices of $\alpha_{\rm NUV}\approx-1$ and
  $\alpha_{\rm EUV}\approx-2$ at $\lambda \gapp 1200$~\AA\ and $\lambda \lapp 1200$~\AA, respectively (see
  \citealt{ssd12} and references therein).}
However, if we assume this to be valid, our model \citep{cw12}, which applies the equation of photoionsation
equilibrium \citep{ost89} to an exponential gas disk, yields a scale-length of $R=2.8$~kpc for $Q_\text{\HI}=4.8
\times10^{56}$~sec$^{-1}$, using the canonical gas temperature of 2000~K, or $R=2.1$~kpc, using the $T=500$~K typical of
intervening systems \citep{cras16,cur17a}. This compares to $3.15$ kpc for the Milky Way \citep{kk09}, showing that an
ionising rate of $Q_\text{\HI}=4.8 \times10^{56}$~sec$^{-1}$ is insufficient to completely ionise the gas in a large spiral,
thus not requiring a re-evaluation of the critical ionising photon rate.
%papers/59/X-n-Mcyl_r_overlay.c    

Of the 512 sources with sufficient UV photometry, below the highest value where 21-cm absorption has been detected there
are 106 detections and 406 non-detections, giving a detection rate of 20.7\%. Using a detection probability of $p=0.207$, where
$p=0.5$ is even odds, the binomial probability of obtaining the observed zero detections out of 50 searches for 21-cm
absorption is $P({\rm bin}) = 9.18\times10^{-6}$. Assuming Gaussian statistics, this is significant at $S({\rm bin}) =
4.42\sigma$. This is lower than the $6.83\sigma$ significance reported by \citet{chj+17}, although only $z\geq0.1$
sources were considered and below these redshifts there may be another factor at play which lowers the detection rate
(see Sect. \ref{lse}).

It should be noted that, while there are 48 quasars above the critical ionising photon rate, there are only two
galaxies.  It is therefore tempting to attribute the lack of detection of 21-cm absorption above the critical rate to an
unfavourable alignment of the continuum emission and absorbing gas along our sight-line. However, although all of the
sources with $L_{\rm UV}\gapp10^{23}$ \WpHz\ are type-1 AGN, for those below this there is a 50\% detection rate for
{\em both} type-1 and type-2 objects \citep{cw10}.\footnote{Based upon 21 type-1 and 33 type-2 objects with $L_{\rm
    UV}<10^{23}$ \WpHz.  For the current sample there is a 21.8\% detection rate for galaxies ($n=428$) and 14.3\% for
  quasars ($n=132$) at $Q_\text{\HI}>4.8 \times10^{56}$~sec$^{-1}$.}
% awk < low_G.txt '{if ($5 ~ /det/) print $0}' | wc -l      93
%awk < low_G.txt '{if ($5 ~ /non/) print $0}' | wc -l    333    => 21.8% 
% awk < low_Q.txt '{if ($5 ~ /det/) print $0}' | wc -l      12
% awk < low_Q.txt '{if ($5 ~ /non/) print $0}' | wc -l     72     => 14.3%
Thus, the orientation of inner torus has little bearing upon the detection of 21-cm absorption, which must occur in a
randomly oriented large-scale galactic disk (e.g. \citealt{nw99}).\footnote{Although a small contribution from
the torus is detected in the stacked absorption spectra \citep{cdda16}.} 

There does, however, remain the possibility that in all of the $L_{\rm UV}\gapp10^{23}$ \WpHz\ objects the gas simply
does not occult the unobscured AGN along our sight-line. This would imply a selection effect where the higher UV flux is
due to the absence of dense intervening gas, representing a sub-sample where the torus and disk are roughly aligned,
thus differing from the $L_{\rm UV}\lapp10^{23}$ \WpHz\ type-1 objects. However, the complete ionisation of the neutral
atomic gas in a large galaxy at this luminosity is consistent with the model of \citet{cw12}. Furthermore,
\citet{psv+12} note a critical X-ray luminosity, above which sources are not detected in 250~$\mu$m emission, which
is a tracer of star formation. The fact that it is not detected in {\em emission} above a certain
value cannot be attributed to orientation. Whether due to  complete ionisation of the gas or a selection bias towards type-1
objects, where the galactic disk is coplanar with the torus, the fact remains that \HI\ 21-cm has not been detected above
$Q_\text{\HI}\sim5\times10^{56}$~sec$^{-1}$.

\subsection{Radio} 

\HI\ 21-cm absorption traces the cool component of the neutral gas and, in addition to ionisation of \HI\ by $\lambda
\leq912$ \AA\ photons, the non-detection of 21-cm absorption can be caused by excitation to
the upper hyperfine level by $\lambda =21$~cm photons \citep{pf56,be69}. Although we have previously ruled this out as a
reason for the non-detections at high redshift \citep{cww+08,cw10,cwt+12}, from a sample of 52 flat-spectrum sources,
\citet{akk16} suggest that the radio luminosity may be as important as the UV luminosity in the detection of 21-cm
absorption. Being ``uniformly selected flat-spectrum sources'' (although the spectral indices range
$-1.2\lapp\alpha\lapp1.2$), these form a biased sub-sample of redshifted radio sources.

\begin{figure*}
\centering 
\includegraphics[angle=270,scale=0.59]{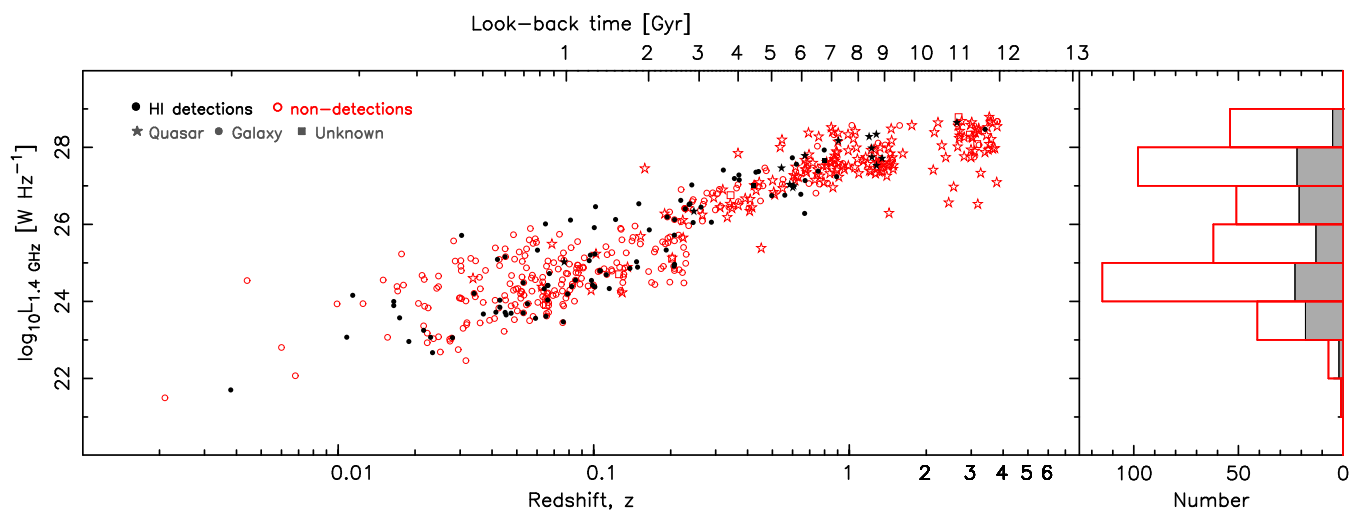}
\caption{The rest-frame 1.4 GHz continuum luminosity versus redshift for the \HI\ 21-cm absorption searches. As per
  Fig. \ref{Q-z}, the filled symbols/histogram represent the detections and the unfilled symbols/histogram the
  non-detections.}
\label{L21-z}
\end{figure*}

In Fig.~\ref{L21-z},  we show the distribution of 1.4~GHz continuum luminosities, 
where the highest luminosities represented by the histogram show a mix of detections and non-detections
(cf. Fig.~\ref{Q-z}).  In this case, the highest luminosity at which 21-cm absorption is detected is $L_{1.4~{\rm GHz}} =
4.38\times10^{28}$ \WpHz. Below this value there are 105 detections and 422 non-detections, giving a detection rate of
19.9\%. Applying $p=0.199$ to the sources with radio luminosities of $L_{1.4~{\rm GHz}} >4.38\times10^{28}$ \WpHz,
the binomial probability of zero detections out of nine searches is $P({\rm bin}) =0.135$. which is significant at just
$S({\rm bin}) =1.49\sigma$. Thus, it is clear that the radio luminosity is not as important in the detection of 21-cm absorption and that
any bias towards non-detections at high radio luminosities is probably due to the fact that these will tend also be the
high UV luminosity sources \citep{cw10}.

Addressing the suggestion by \citet{akk16} and \citet{ak17} that the non-detections at high redshift may be due to an unspecified
dependence on redshift, the highest redshift detection is at $z=3.398$ \citep{ubc91}, below which there are 105
detections and 416 non-detections (a 20.2\% detection rate).  Above this redshift there are 19 non-detections, which
gives a binomial probability of $P({\rm bin}) =0.034$, which is significant at $S({\rm bin}) =2.12\sigma$. Again, this is not as significant as the
effect of the UV luminosity, although certainly degenerate with it. 

Statistically, it is therefore clear that the ionisation of the neutral gas is the dominant effect in the non-detection
of associated 21-cm absorption at high redshift. The value of the critical luminosity also remains consistent with the
model of \citet{cw12}, which shows that rates of $Q_\text{\HI}\gapp5\times10^{56}$~sec$^{-1}$ ($L_{\rm
  UV}\gapp3\times10^{23}$ \WpHz) are required to fully ionise the disk of a large spiral galaxy.

\section{Evolution of the neutral gas}
\label{eng}
\subsection{Line-strength}
\label{lse}

\begin{figure}
\centering 
\includegraphics[angle=270,scale=0.52]{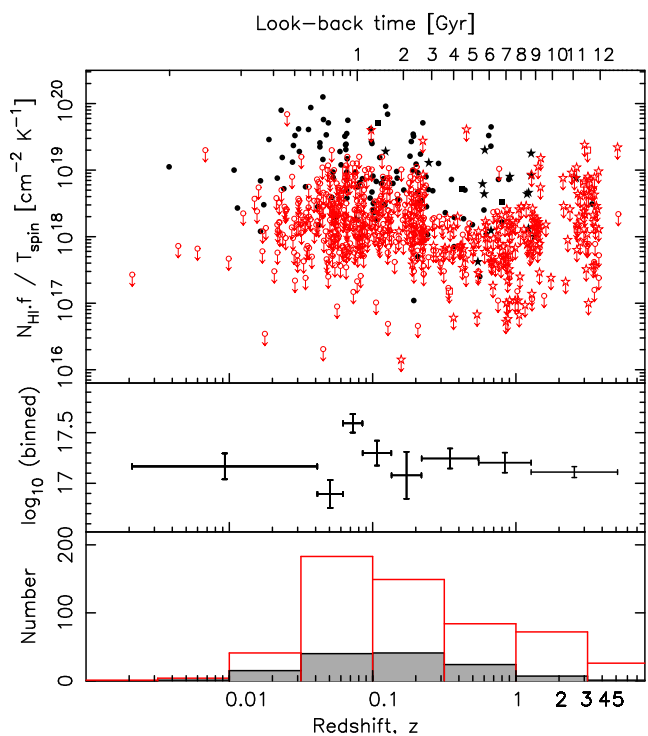}
\caption{The line strength ($1.823\times10^{18}\int\!\tau_{\rm obs}\,dv$) versus redshift for the associated
  \HI\ 21-cm absorption searches. The filled symbols/histogram represent the detections and the unfilled
  symbols/histogram the $3\sigma$ upper limits to the non-detections. The middle panel shows the binned values in
  equally sized bins, including the limits.  The horizontal error
  bars show the range of points in the bin and the vertical error bars the error  in the mean value. In
  the bottom panel the filled histogram shows the number of detections and the unfilled the non-detections.}
\label{N-z}
\end{figure}

As mentioned in Sect. \ref{intro}, there is evidence that the \HI\ 21-cm absorption strength in intervening systems 
may trace the star formation history of the Universe \citep{cur17,cur17a}.  Plotting the 21-cm absorption strength
versus redshift, we also see a peak for the associated systems (Fig. \ref{N-z})\footnote{As described in \citet{chj+17},
  all of the limits are re-sampled to the same spectral resolution of 20 \kms. For the binning in the middle panel, the
  upper limits are included as censored data points, via the {\em Astronomy SURVival Analysis} ({\sc asurv}) package
  \citep{ifn86}. The points are binned via the Kaplan--Meier estimator, giving the maximum-likelihood estimate based
  upon the parent population \citep{fn85}.}, although this is at a redshift of $z\sim0.1$, cf. $z\sim2.5$ for the
intervening absorbers.
The decrease in strength with increasing redshift can be attributed to increasing UV luminosities (Sect. \ref{ciul}), at
least at $z\gapp0.1$, where there is a $3.66\sigma$ anti-correlation between $N_\text{\HI}.f/T_{\rm spin}$ and
$Q_\text{\HI}$ \citep{cwa+17}. Below $z\sim0.1$, however, the cause is not clear, although the numbers are small (at
least at $z\lapp0.05$, Fig. \ref{N-z}, bottom panel). A likely possibility is that at these redshifts 21-cm emission
becomes more detectable and the decrease in the absorption line strength may be due to dilution by 21-cm emission
(e.g. \citealt{rsa+15}).

\subsection{Spin temperature}

When combined with the total neutral hydrogen column density, $N_\text{\HI}$, the spin temperature,
$T_{\rm spin}$, degenerate with the covering factor $f$, may be estimated. Assuming that the Lyman-\AL\ absorption,
which yields $N_\text{\HI}$ [\scm], and the 21-cm absorption arise in the same sight-line, 
in the optically thin case (when the observed optical depth is $\tau_{\rm obs}\lapp0.3$), 
$T_{\rm spin}$ [K] is given by \citep{wb75}
\[
\frac{T_{\rm  spin}}{f} = \frac{N_\text{\HI}}{1.823\times10^{18}\int\!\tau_{\rm obs}\,dv},
\]
where $\int\!\tau_{\rm obs}\,dv$ [\kms] is the observed velocity
integrated optical depth of the 21-cm absorption.  For the associated absorbers, the column densities are generally not
available but, using the method of \citet{cur17a}, if we assume that the associated absorbers follow the same evolution
of the cosmological mass density as the intervening absorbers, then from $\Omega_{\text{\HI}} = 4.0\times10^{-4}(z_{\rm
  abs}+1)^{0.60}$ \citep{cmp+17} and the redshift number density of damped Lyman-\AL\ absorption systems $n_{\rm DLA} =
0.027(z+1)^{1.682}$ \citep{rtsm17} 
\[
\left< N_\text{\HI} \right> = 0.011(z+1)^{0.92} \frac{3\,H_{0}^2}{8\,\pi\,G}\frac{c}{m_{_{\rm H}} H_{\rm z}},
\]
where $G$ is the gravitational constant, $c$ is the speed of light, $m_{_{\rm H}}$ is the mass of the hydrogen atom
and $H_{\rm z}$ is the Hubble parameter at redshift $z$.

\begin{figure}
\centering 
\includegraphics[angle=270,scale=0.52]{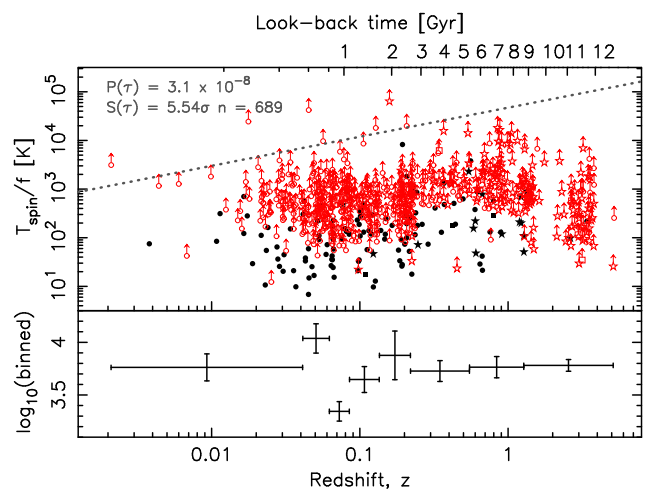}
\caption{The estimated spin temperature (degenerate with the covering factor) for the associated 21-cm absorption
  searches.  $P(\tau)$ shows the probability of the correlation arising by chance and $S(\tau)$ the associated
  significance, from a generalised non-parametric Kendall-tau test. The dotted line shows the linear regression by
  the Buckley--James method within the {\sc asurv} package, which includes the lower limits. The symbols and error bars 
in the top and bottom panels, respectively, 
are as per Fig. \ref{N-z}.}
\label{normN}
\end{figure}

Applying the derived $\left< N_\text{\HI}\right>-z$ relation to the associated absorbers, gives the distribution shown
in Fig. \ref{normN},
where we essentially see an inverted version of the line strength distribution (Fig.~\ref{N-z}), due to the generally
flat evolution of $\left< N_\text{\HI}\right>$. As per the intervening absorbers, the degeneracy with the covering
factor is difficult to break, although, since the absorption occurs at the same redshift as the continuum emission, there
will be no bias due to the absorber and emitter being at different angular diameter distances \citep{cw06}. Since the
covering factor can have a value $\tau_{\rm obs} <f \leq1$ (\citealt{obg94}), the values in Fig.~\ref{normN} represent lower
limits to the spin temperature for all of the measurements.

If we take this at face value, however, there is an apparent increase in the spin temperature with redshift, which 
may be expected from the selection of more UV luminous sources.  The lowest value for which 
$Q_\text{\HI}>4.8\times10^{56}$~sec$^{-1}$ is $T_{\rm spin}/f \geq 30$~K, not a particularly tight constraint, or even
applicable, given that we expect  complete ionisation of the neutral gas. On the other hand, 
the highest value at which 21-cm absorption has been detected is $T_{\rm spin}/f =8270$~K, similar to that of the
intervening absorbers \citep{cur17}. 

\section{WISE colours}
\label{nic}
\subsection{Redshift evolution}
\label{gqc}

\begin{figure}
\centering \includegraphics[angle=-90,scale=0.52]{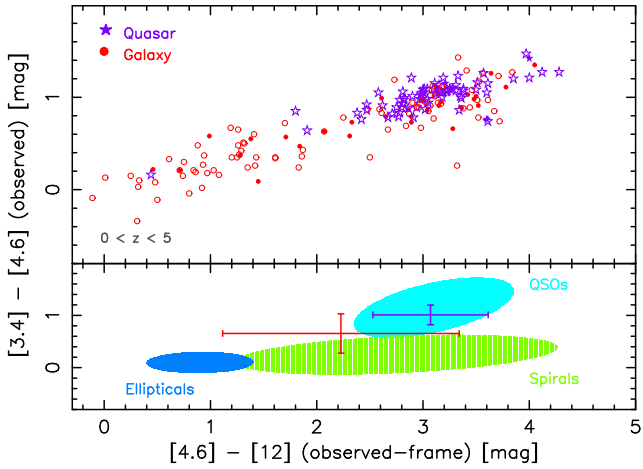}
\caption{The observed-frame WISE colours for the sample, where available. Again, the filled symbols indicate 21-cm
  detections and the unfilled symbols non-detections, with the shapes representing the classification (stars--quasars,
  circles--galaxies, squares--unknown).  The bottom panel shows the binned values, where the error bars show
  $\pm1\sigma$ and the overlays show the colours of object classes \citep{wem+10,gas+17}.}
\label{WISE_obs}
\end{figure}

The mid-infrared colours can provide insight into the activity within a source \citep{wem+10}, specifically the bands
observed by the Wide-Field Infrared Survey Explorer, where the $\lambda=3.4$~$\mu$m (WISE W1) and 4.6 $\mu$m (WISE W2) bands
trace hot dust in the sub-pc torus (by AGN activity) and the 12 $\mu$m (WISE W3) band traces the dust heated in the host
galaxy \citep{jcm+11,dyt+12}. In Fig.~\ref{WISE_obs}, we show the observed-frame WISE colours for the sources searched
in \HI\ 21-cm absorption, where available.
From this, we see a separation between the galaxies and quasars as expected from their WISE colours \citep{wem+10},
although the (radio) galaxies have elevated $[3.4] - [4.6]$ (WISE W1--W2)  colours in relation to the expected distribution
of spirals.

\begin{figure}
\centering \includegraphics[angle=-90,scale=0.52]{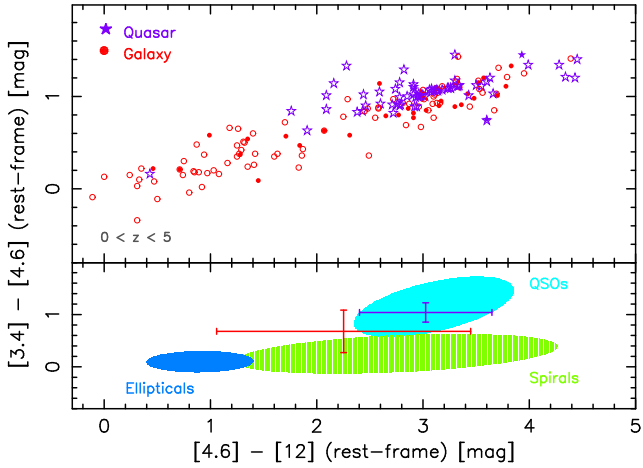}
\caption{As Fig. \ref{WISE_obs}, but in the rest-frame of the source.}
\label{WISE_rest}
\end{figure}

\begin{figure}
\centering \includegraphics[angle=-90,scale=0.52]{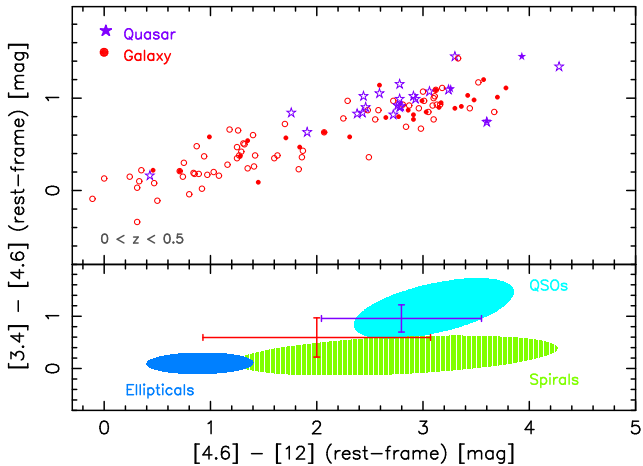}
\caption{As Fig. \ref{WISE_rest}, but for $z < 0.5$ sources only.}
\label{WISE-low}
\end{figure}

\begin{figure}
\centering \includegraphics[angle=-90,scale=0.52]{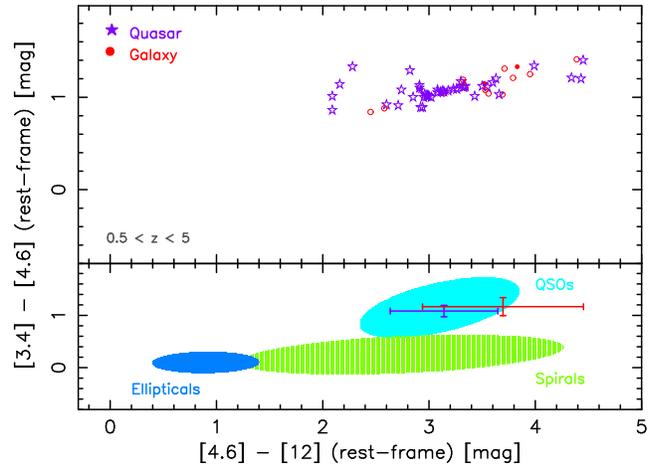}
\caption{As Fig. \ref{WISE_rest}, but for $z\geq 0.5$ sources only.}
\label{WISE-high}
\end{figure}

Since the  sources span a large range of redshifts, $0 \lapp z \lapp 5$,  cf. $z<0.10$ by \citet{gas+17} and $z<0.23$ by \citet{cs17}, 
like \citet{cjh+14}, we also show the colours where the infrared magnitudes have been obtained from the
rest-frame SED (Fig. \ref{WISE_rest}). In addition to this, 
in Figs. \ref{WISE-low} and \ref{WISE-high} we split the sample into low ($z < 0.5$) 
and high ($z\geq 0.5$) redshift bins.
The low redshift sample shows a similar distribution to the previous studies (\citealt{cs17,gas+17}),
with the mean galaxy colour overlapping, although off-centred, to that expected for spirals \citep{wem+10}.
For the high redshift sample, however, both classes move up in $[4.6] - [12]$, indicating
more vigorous star formation. In addition to this, the galaxies move up significantly in $[3.4] - [4.6]$, 
becoming more ``quasar-like''.

\begin{figure}
\centering \includegraphics[angle=-90,scale=0.55]{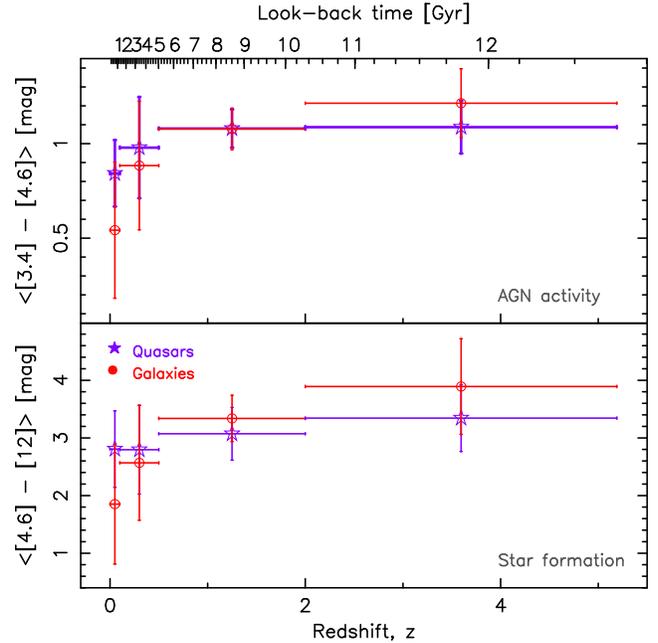}
\caption{The WISE colours versus redshift. The vertical error bars show $\pm1\sigma$ in the mean value and
  the horizontal error bars show the range of points.}
\label{colour_evolv}
\end{figure}

To better illustrate these changes, in Fig. \ref{colour_evolv} we show how the colours of both classes evolve with redshift,
from which we see that galaxies may increase both their star forming and AGN activity more rapidly than the quasars,
although the classes are consistent to within $1\sigma$. A possible explanation is that only the brightest objects are
detected at high redshift. Although this constitutes a possible selection effect, it does tell us that previous searches
for high redshift 21-cm absorption, which rely upon an optical redshift, are biased towards different objects than at
low redshift (Sect. \ref{ciul}).

Finally, the redshift evolution of the colours indicates that the morphologies of high redshift galaxies searched for
21-cm absorption are unlikely to be dominated by ellipticals (Fig. \ref{WISE-high}). This is consistent with the
hypothesis of \citet{cw10}, that photo-ionisation, rather than a selections bias towards these gas-poor galaxies, is the
cause of the low detection rates.

\subsection{Colour--{\boldmath $z$} relations}
\label{C-z}

From the above colour dependences on redshift, it is possible that the WISE colours may provide an estimate of a
redshift where an optical spectrum is not available or desired. For example, as a means of estimating redshifts for the
large number of sources expected to be detected through forthcoming radio continuum surveys (e.g. the Evolutionary Map
of the Universe, \citealt{nha+11}), which would otherwise be observationally expensive. Also, in the search for high
redshift atomic and molecular gas in absorption, where optical brightness can select
against a detection (\citealt{cww+08} \&  \citealt{cwc+11}, respectively, see Sect. \ref{ciul}).

\begin{figure}
\centering \includegraphics[angle=-90,scale=0.52]{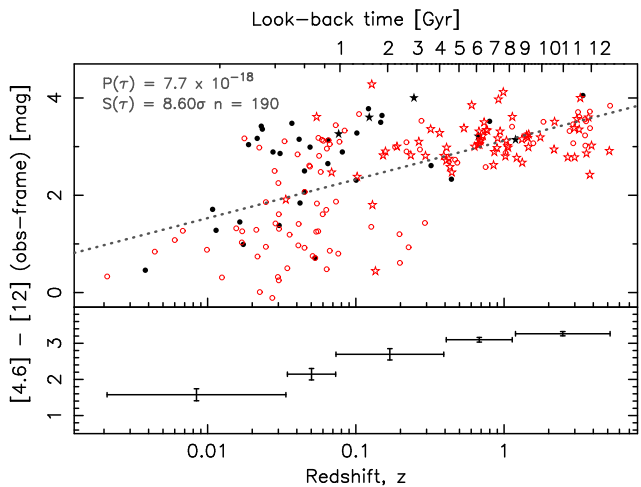}
\caption{The $[4.6] - [12]$ observed-fame colour versus redshift. $P(\tau)$ shows the probability of the correlation
  arising by chance and $S(\tau)$ the associated significance, from a generalised non-parametric Kendall-tau test.  The
  dotted line shows the best fit (see Table \ref{z_stats}). As previously, the filled symbols represent the 21-cm
  detections and the unfilled symbols the non-detections, with the shapes representing the classification
  (stars--quasars, circles--galaxies, squares--unknown).}
\label{W2-W3_z}
\end{figure}

\begin{figure}
\centering \includegraphics[angle=-90,scale=0.52]{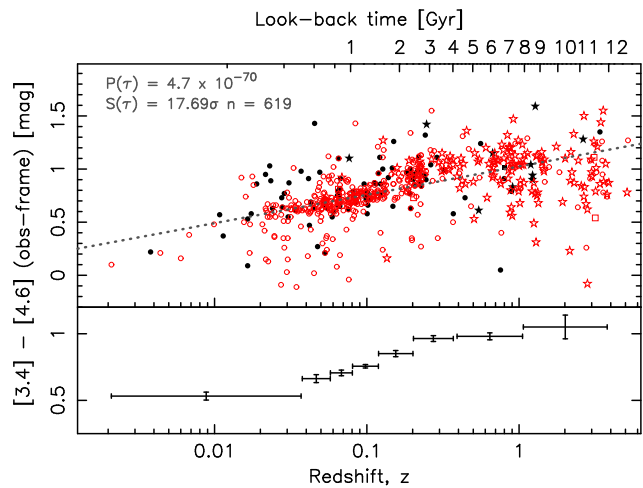}
\caption{As Fig. \ref{W2-W3_z}, but for the $[3.4] - [4.6]$ colour.}
\label{W1-W2_z}
\end{figure}

A correlation between the $\lambda=2.2$~$\mu$m $K$-band magnitude and redshift is well documented
\citep{ll82,erl+97,jre+01,dvs+02,wrjb03,itm+10}, as well as for the near-by 3.4~$\mu$m (WISE W1) band
\citep{cbn+14,gas+18}. In Figs. \ref{W2-W3_z} and \ref{W1-W2_z} we show the $[4.6] - [12]$ and $[3.4] - [4.6]$ colours versus redshift,
which are very strongly correlated [$S(\tau)=8.60\sigma$ and $17.69\sigma$, respectively].
The stronger correlation for the $[3.4] - [4.6]$ colour  could be due to the larger sample and
so, in order to compare the correlations on an equal footing, we perform a Monte Carlo simulation with 
10\,000 trials, where for each we select 190 $[3.4] - [4.6]$--$z$ pairs at random from the 619 available. 
\begin{figure}
\centering \includegraphics[angle=-90,scale=0.50]{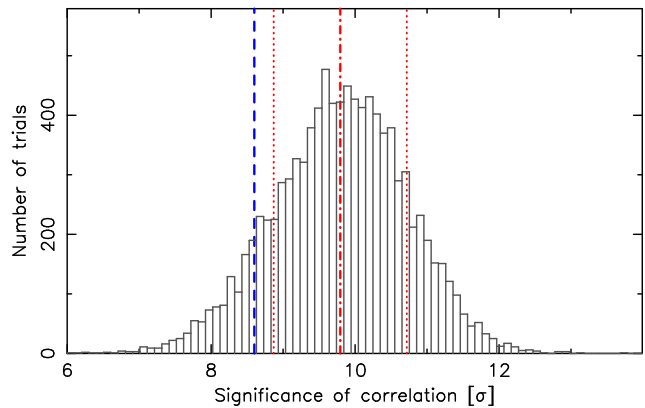}
\caption{The distribution of the significance of the correlation for 10\,000 trials of 190 $[3.4] - [4.6]$--$z$ pairs (Fig. \ref{W1-W2_z}).
 The dot-dashed line shows the mean ($\mu=9.794$) and the dotted lines the standard deviation ($\sigma= 0.924$) from this. 
The dashed line shows the significance of the $[4.6] - [12]$--$z$  correlation (Fig. \ref{W2-W3_z}).}
\label{rand_z}
\end{figure}
Showing the results in Fig.~\ref{rand_z}, it does appear possible that the $[4.6] - [12]$--$z$ correlation may be just as significant.

We can compare each parameter's effectiveness as a redshift estimator via machine learning techniques, using parameters
which do not rely upon knowing the redshift of the source. We select the observed 1.4 GHz flux, $S_{\rm obs}$, the radio-band
turnover frequency, $\nu_{\rm TO}$, the radio-band spectral index, $\alpha$, and the blue and mid-infrared magnitudes/colours
as the {\em features} for the {\sc weka} package \citep{hfh+09}. We create several classes by dividing the redshift
space into $\Delta\log_{10}(z+1) =0.1$ bins, from which  we obtain an accuracy of
$68.5-71.4$\% (depending upon the algorithm, see \citealt{cdda16}) in predicting the correct class. 
In Table \ref{z_stats}, we summarise the significance and fits of the features in order of descending ranking.
\begin{table*} 
\centering
\begin{minipage}{145mm}
  \caption{Parameter fits to $\log_{10}z$ in order of decreasing importance, given by the Pearson's correlation
    coefficient, $r$, for $\Delta\log_{10}(z+1) =0.1$ wide classes, using the Random Forest algorithm \citep{hfh+09}.
    The magnitudes and colours are in the observed fame.  $P(\tau)$ gives the probability of the correlation arising by
    chance, from a generalised non-parametric Kendall-tau test, and $S(\tau)$ the associated significance. Note that,
    for the sake of comparison, for the fluxes and magnitudes we assume the null hypothesis that these are independent
    of redshift.  The gradient and intercept are obtained via linear regression by the Buckley--James
    method.}
\begin{tabular}{@{}l l c  c c  r r c c  r  @{}} 
\hline
\smallskip
Feature &  & $n$ & $r$ & $P(\tau)$ & $S(\tau)$ & Gradient, $m$ & $\Delta m/m$ & $\sigma_{m}$ & Intercept \\
\hline
$\log_{10} S_{\rm obs}$  & & 686 & 0.3177 & 0.0675 & $1.83\sigma$ & $-0.034\pm0.063$  & 1.85 & 1.05 & 0.16\\ 
3.4 $\mu$m mag & & 618 &0.1889  & $\approx0$&  $23.08\sigma$ & $2.194\pm0.084$ &  0.04 & 1.31 & 13.92 \\
 4.6 $\mu$m mag & &  618 & 0.1706  &   $\approx0$     & $20.88\sigma$ & $2.017\pm0.082$ &  0.04 & 1.28 & 13.02 \\
$K$  magnitude  & & 510 &0.1667   & $\approx0$ & $24.23\sigma$ & $2.604\pm0.070$ & 0.03 & 0.97 & 14.98\\ 
$B$  magnitude  && 565 &   0.1368 & $7.21\times10^{-90}$& $20.01\sigma$ & $1.775\pm0.251$ & 0.14 & 3.82 & 18.85\\ 
$[3.4] - [4.6]$ & & 619 & 0.1200& $4.75\times10^{-70}$ & $17.69\sigma$ & $0.267\pm0.022$ & 0.08 & 0.34 & 1.03\\ 
$B-K$     & & 572  &   0.0600 & $2.10\times10^{-5}$ & $4.25\sigma$ & $-0.469\pm0.076$ & 0.16 & 1.08 & 4.54\\ 
Spectral index, $\alpha$ & & 576 &  0.0450 & $3.36\times10^{-9}$ & $5.91\sigma$ & $0.185\pm0.038$ & 0.21 & 0.62 & $-0.20$\\ 
12 $\mu$m mag & &190  &   0.0361 & $7.94\times10^{-16}$ & $8.06\sigma$ & $1.76\pm0.18$ & 0.10& 1.97 & 10.15\\ 
$[4.6] - [12]$ & & 190 &   0.0195& $7.69\times10^{-18}$ & $8.60\sigma$ & $0.796\pm0.073$ &  0.09& 0.80 & 3.12\\
$\nu_{\rm TO}$         &    TO            & 176 & 0 & 0.049 & $1.97\sigma$ & $0.221\pm0.111$ &  0.50& 0.95 & 9.71\\
& All$^{\dagger}$ & 369 & --- &  0.078 & $1.76\sigma$ & $0.372\pm0.111$ &  0.30& 0.95 & 8.54\\
   \hline
 \end{tabular}
 {$^{\dagger}$Assumes that where a turnover in the radio SED has not been detected that $\nu_{\rm TO}$ is located below the lowest observed
   frequency and so is included as an upper limit (see \citealt{cwa+17}). The previous row is for where a turnover is detected only, which, unlike the limits,  can
   be readily input to  {\sc weka}.}
\label{z_stats} 
\end{minipage}
\end{table*} 
This confirms the finding of \citet{gas+18} that the 3.4 $\mu$m and 4.6 $\mu$m magnitudes are more effective than most
other indicators, although, not surprisingly, similar to the $K$-band (2.2 $\mu$m) effectiveness.  All values, however,
have a low correlation coefficient $|r| \ll 1$, with the WISE colours not being highly ranked, although the
redshift fit could be ``tighter'' than from other features, since the uncertainties in the fitted slope are
similar to that of the near--infrared magnitudes ($\Delta m/m \leq 0.1$) and so these may be the key
to the prediction of a statistical redshift:
While significant correlations between the
radio spectral index and redshift have previously been noted (e.g. \citealt{ak98,dvs+02}), the scatter is generally too
large to permit a useful prediction \citep{maj15}.

\subsection{\HI\ detection}

Although not seen in their own sample of compact ($\theta_{\text{20 GHz}}<15$ arcsec) radio sources, \citet{gas+17} note
that all of the flux selected ($S_{\text{1.4 GHz}} > 50$ mJy) sources of \citet{gmmo14}, with WISE colours of $[4.6] -
[12] >2.7$, are detected in 21-cm absorption.
Furthermore, \citet{cs17} note a high detection rate (70\%) in compact objects with $[4.6] - [12] >2.0$.  
Since, for the whole sample, the $[4.6] - [12]$ values increase with redshift (Sect.~\ref{gqc}), we examine the
  binomial probability of the number of detections or more about the median value (Table~\ref{W2-W3_stats}), which
  exceeds $2.7$ above $z=0.5$ (Fig.~\ref{WISE-high}).
\begin{table} 
\centering
\caption{The $[4.6] - [12]$ (WISE W2--W3) \HI\ 21-cm detection statistics for $n$ values within the redshift range. The
  third column gives the median $[4.6] - [12]$ colour of the quoted redshift range (Fig.~\ref{colour_evolv}), followed
  by the detection rate statistics below and above this.  The last two columns give the binomial probability and
  significance of this observed number of detections or more at $[4.6] - [12]\geq$\, the median, given the detection
  rate below the median value.}
\begin{tabular}{@{}l c  c c  c c  c  @{}} 
\hline
\smallskip
$z$-range & $n$  & Median & \multicolumn{2}{c}{Rate [\%]} & $P({\rm bin})$ & $S({\rm bin})$\\ 
                   &             &      Below   &  Above &  & \\
  \hline
 All             &   173 &    2.82      &      16.3    &  25.3   &       0.021          &     $2.31\sigma$          \\
 $<0.1 $      &   82 &  1.84     &          17.1      &  39.0   &    $7.04\times10^{-4}$        &    $3.39\sigma$          \\ 
$0.1- 0.5$   &  39  & 2.78      &         17.6      &  27.3   &  0.179   & $1.34\sigma$  \\
$0.5-2$        & 36  & 3.11    &       5.6            & 11.1 &   0.264 &   $1.12\sigma$ \\
$2-5$            & 15 & 3.55   &       0                & 12.5 & --- &  ---\\
\hline
\end{tabular}
\label{W2-W3_stats}  
\end{table} 
% awk < all-info.dat '{if ($8~/gmmo14/) print $5}' > gmmo_z.dat     Expectation value =  0.1054
% awk < all-info.dat '{if ($8 ~/gas/) print $5}' > gas+17_z.dat    Expectation value =  0.06626
From the table, we only see a correlation at the lowest redshifts which is consistent with that found for 
$z\leq0.23$ sources by \citet{cs17} and in the \citet{gmmo14} data.

\citet{gas+17} suggest that there is also an anti-correlation between the $[3.4] - [4.6]$ colour and the 21-cm detection rate,
which could be consistent with excitation of the neutral gas by the AGN (Sect. \ref{epf}). However, any
such correlation is weak (Table~\ref{W1-W2_stats}).
\begin{table} 
\centering
\caption{As Table \ref{W2-W3_stats}, but for the $[3.4] - [4.6]$ (WISE W1--W2)  colours.}
\begin{tabular}{@{}l c c   c  c c  c  @{}} 
\hline
\smallskip
$z$-range & $n$ & Median & \multicolumn{2}{c}{Rate [\%]} & $P({\rm bin})$ & $S({\rm bin})$\\ 
                   &        &      &      Below   &  Above &  & \\
  \hline
 All             &    526 &  0.75      &    16.1    &  20.2   &   0.056 &  $1.91\sigma$   \\
$<0.1 $      &   273 &  0.67   &     17.6      &  19.7   &       0.280        &    $1.08\sigma$          \\ 
$0.1- 0.5$   &  196  & 0.88      &    18.4    &  23.5 &  0.122  & $1.55\sigma$  \\
$0.5-2$        &   39& 1.07      &       5.3     & 10.0&  0.284&  $1.07\sigma$\\
$2-5$            &   & 1.16   &   0     & 11.1 & --- & ---\\
\hline
\end{tabular}
\label{W1-W2_stats}  
\end{table} 

\subsection{Photo-ionisation}
\label{p-i}

\begin{figure}
\centering \includegraphics[angle=-90,scale=0.52]{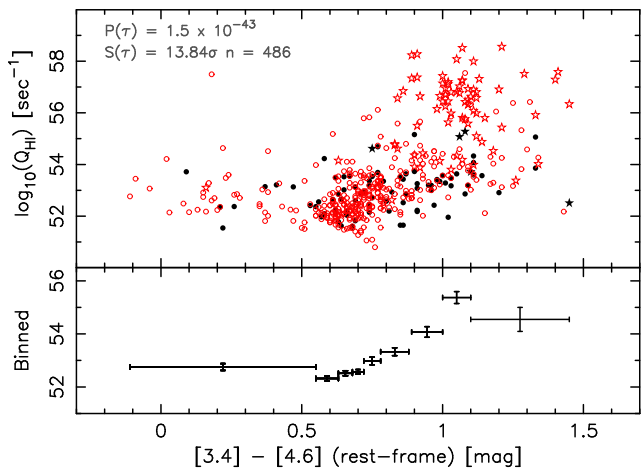}
\caption{The ionising ($\lambda \leq912$ \AA) photon rate versus the rest-frame $[3.4] - [4.6]$ colour.  As above, the
  filled symbols represent the 21-cm detections and the unfilled symbols the non-detections, with the shapes representing the
  classification (stars--quasars, circles--galaxies, squares -- unknown). The error bars are as per Fig. \ref{N-z}.}
\label{W1-W2_UV}
\end{figure}

\begin{figure}
\centering \includegraphics[angle=-90,scale=0.52]{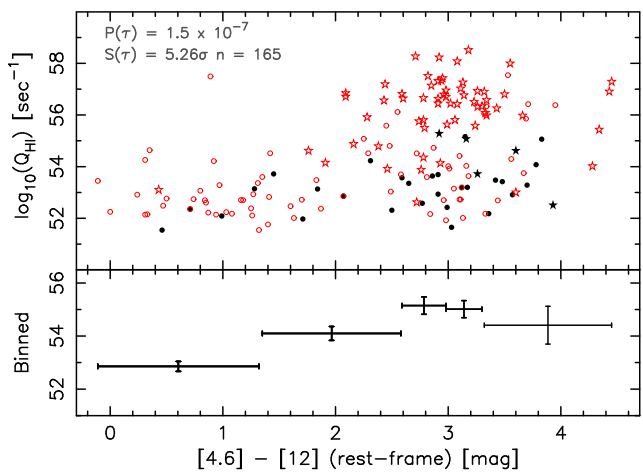}
\caption{As per Fig. \ref{W1-W2_UV}, but for the rest-frame $[4.6] - [12]$ colour.}
\label{W2-W3_UV}
\end{figure}

\begin{figure}
\centering \includegraphics[angle=-90,scale=0.50]{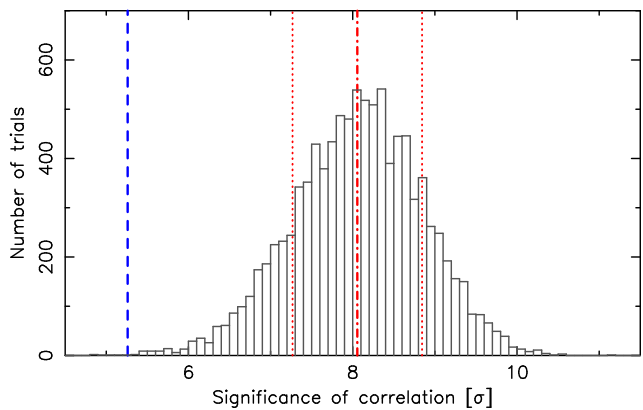}
\caption{The distribution of the significance of the correlation for 10\,000 trials of 165 $Q_\text{\HI}$--$[3.4] - [4.6]$ pairs.
 The dot-dashed line shows the mean ($\mu=8.056$) and the dotted lines the standard deviation ($\sigma= 0.788$) from this. 
The dashed line shows the significance of the $Q_\text{\HI}$---$[4.6] - [12]$ correlation (Fig. \ref{W2-W3_UV}).}
\label{rand}
\end{figure}

Although the AGN activity, as traced through the $[3.4] - [4.6]$ colour, appears to have little effect on the
21-cm detection rate (Table \ref{W1-W2_stats}), we know that the UV luminosity does. From the ionising photon rate
versus the $[3.4] - [4.6]$ colour (Fig. \ref{W1-W2_UV}),
we find a very strong correlation [$S(\tau)=13.84\sigma$]. Of course, it is possible that some of the UV flux is due to the presence of young
stars (see \citealt{caw+16}), and so we also consider the relation between the $[4.6] - [12]$ colours (Fig.~\ref{W2-W3_UV}).
Although significant, this does not appear to be as strongly correlated as the $[3.4] - [4.6]$ colour. However, this
only consists of 165 sources for which both $[4.6] - [12]$ and $Q_\text{\HI}$ can be obtained, cf. 486 sources for
$[3.4] - [4.6]$.  Therefore, as in Sect. \ref{C-z}, we compare the correlations by selecting 165 $Q_\text{\HI}$---$[3.4]
- [4.6]$ pairs at random from the 486 available, the results of which are shown in Fig.~\ref{rand}.
From this, it is clear that the $Q_\text{\HI}$---$[3.4] - [4.6]$ correlation is stronger than that of $Q_\text{\HI}$---$[4.6] -
[12]$, suggesting a dominant  AGN contribution to the UV flux.

\section{Summary}
\label{sum}

Upon updating the redshifted \HI\ 21-cm absorption catalogue with the addition of $z\lapp0.1$ sources,
we revisit the effects of the radio and ultra-violet luminosities, as well as investigating the effects of the
mid-infrared colours, on the detection of cool, neutral gas in the distant Universe. We find:
\begin{enumerate}
\item For the ultra-violet and radio luminosities:
 \begin{enumerate}
\item That the ionising photon rate remains the dominant factor, over and above radio luminosity and
   redshift, in the detection of 21-cm absorption in the host galaxy. 
  \item The highest ionising photon rate at which 21-cm absorption has been detected may be
     $Q_\text{\HI}=5\times10^{56}$~sec$^{-1}$ ($L_{\rm UV}\approx 3\times10^{23}$ \WpHz). This is barely sufficient to
     completely ionise the gas in the large spiral, remaining consistent with the model of \citet{cw12}.
\item While the increase in ultra-violet luminosity is responsible for a decreasing 21-cm detection rate at $z\gapp0.1$ \citep{cw10},
  we also note a decrease at  low redshift. We suggest that this is due to increased ``contamination'' by 21-cm emission, which
strengthens with decreasing redshift, thus diluting the absorption strength.
         \end{enumerate}
\item For the mid-infrared colours:
\begin{enumerate}
 \item While the $[4.6] - [12]$ colours increase for both quasars and galaxies with redshift, the increase in $[3.4] - [4.6]$
   colour is more pronounced for galaxies. Both increases can be attributed to the detection of only the brightest
objects as the redshift increases, although the relative AGN contribution in the high redshift galaxies is larger than at low redshift. This will 
   introduce a bias to optically selected \HI\  21-cm absorption searches in high redshift radio galaxies.
 \item Thus, the colours may be used to predict the source redshift, although the correlations are not as
strong as for the $\lambda=3.4$ and 4.6 $\mu$m (\citealt{cbn+14,gas+18}) nor the $K$-band  ($\lambda=2.2$~$\mu$m)
magnitudes (e.g. \citealt{ll82,dvs+02}).
\item The high redshift sources searched in 21-cm absorption are unlikely to be ellipticals. A possible bias towards the
  selection of these gas-poor objects could offer an alternative explanation to the photo-ionisation being responsible
  for the lack of detections at high redshift. The WISE colours suggest that this is not the case.
\item Both the $[3.4] - [4.6]$ and $[4.6] - [12]$ colours are correlated with the ionsing photon rate, although this
is stronger for $[3.4] - [4.6]$. This indicates that the ultra-violet luminosity is dominated by the AGN, rather than
star-forming, activity, particularly at high redshift. This is consistent with  the argument that star-formation is likely
to be suppressed at $Q_\text{\HI}\gapp10^{56}$~sec$^{-1}$ ($L_{\rm UV}\gapp10^{23}$ \WpHz, \citealt{cw12}).
\item Like other studies, we find a correlation between the 21-cm detection rate with $[4.6] - [12]$ colour, although
  this is only evident at low redshift $z\lapp0.1$. There is no strong correlation with the $[3.4] - [4.6]$ colour.
\end{enumerate}
\end{enumerate}

\section*{Acknowledgements}

We wish to thank James Allison, Stas Shabala, Marcin Glowacki and Liz Mahoney for their input.  
SWD acknowledges receipt of a Victoria Doctoral Scholarship. This research has made
use of {\sc Astropy}, a community-developed core Python package for Astronomy (Astropy Collaboration, 2013), the
NASA/IPAC Extragalactic Database (NED) which is operated by the Jet Propulsion Laboratory, California Institute of
Technology, under contract with the National Aeronautics and Space Administration. This research has also made use of
NASA's Astrophysics Data System Bibliographic Services and {\sc asurv} Rev 1.2 \citep{lif92a}, which implements the
methods presented in \citet{ifn86}.

%\bibliographystyle{../mn2e}  
%\bibliography{aa,ref}

\label{lastpage}

\end{document}